%% file: main.tex
\documentclass[sigconf]{acmart}
\usepackage[utf8]{inputenc}
\renewcommand\footnotetextcopyrightpermission[1]{} 

\usepackage{graphicx}
\usepackage{caption}
\usepackage{subcaption}
\usepackage{bbm}
\usepackage{color}
\usepackage{hyperref}
\usepackage{multirow}
\usepackage{enumitem}
\usepackage{algorithm}
\usepackage{algorithmic}
\usepackage{enumerate}
\usepackage{amsmath}
\usepackage{booktabs}
\usepackage{multirow}
\usepackage{mathrsfs} 
\usepackage{color}

\usepackage[para,online,flushleft]{threeparttable}

\usepackage{soul}   

\acmDOI{}

\acmConference[FATREC'18]{2nd FATREC Workshop on Responsible Recommendation}{October 2018}{Vancouver, Canada}
\acmYear{2018}
\setcopyright{none}

\begin{document}

\title{Personalizing Fairness-aware Re-ranking}

\author{Weiwen Liu}
\affiliation{
    \institution{The Chinese University of Hong Kong}
    \city{Hong Kong}
}\email{wwliu@cse.cuhk.edu.hk}

\author{Robin Burke}
\affiliation{
    \institution{DePaul University}
    \city{Chicago}
    \state{Illinois}
}\email{rburke@cs.depaul.edu}


\begin{abstract}

Personalized recommendation brings about novel challenges in ensuring fairness, especially in scenarios in which users are not the only stakeholders involved in the recommender system. For example, the system may want to ensure that items from different providers have a fair chance of being recommended. To solve this problem, we propose a Fairness-Aware Re-ranking algorithm (FAR) to balance the ranking quality and provider-side fairness. We iteratively generate the ranking list by trading off between accuracy and the coverage of the providers. Although fair treatment of providers is desirable, users may differ in their receptivity to the addition of this type of diversity. Therefore, personalized user tolerance towards provider diversification is incorporated. Experiments are conducted on both synthetic and real-world data. The results show that our proposed re-ranking algorithm can significantly promote fairness with a slight sacrifice in accuracy and can do so while being attentive to individual user differences.

\end{abstract}

\maketitle

\input{body.tex}

\newpage
\bibliographystyle{ACM-Reference-Format}
\bibliography{myref}

\end{document}

%% file: body.tex
\section{Introduction}
Recommender Systems (RSs) have been widely implemented in various fields, e.g., news, music, e-commerce, movies, etc \cite{jannach2010recommender}.  RSs model user preferences and provide personalized recommendations to meet the user's interests or needs. 

In some recommendation scenarios, however, RSs involve more than one stakeholder; systems designed for such scenarios are referred to as Multi-stakeholder Recommender Systems (MRSs). Compared to the previous user-centered systems, an MRS has a multilateral structure and decisions are made to meet the needs of all the stakeholders \cite{soappaper}. For example, Esty \cite{etsy} is an e-commerce platform for shoppers and small-scale artisans, and products are recommended from artisans to shoppers. A recommender system for such a site needs to consider the demands of both buyers and sellers. A typical MRS considers three roles, namely consumers, providers and platform or system \cite{burke_robin_multisided_nodate}. Consumers expect personalized recommendations to meet their interests and needs. Providers offer items to the system and benefit from consumer choices. The system receives items from providers and recommends them to the consumers. The objectives of different stakeholders can be dependent or conflicting, and simply optimizing the objective of a single stakeholder may affect the utility of the others \cite{abdollahpouri_recommender_2017}.  

Most previous recommendation algorithms aim at predicting the user preferences accurately, and ignore the existence of multiple stakeholders and the fairness issues that come along with them \cite{burke_robin_multisided_nodate}. In particular, a system may fail to achieve fairness across multiple providers (P-fairness) by solely maximizing the objective of the consumer. For example, items from established sellers may dominate recommendation lists while newly-arrived sellers receive less attention.

In this paper, we aim at designing a fairness-aware re-ranking algorithm on top of existing recommendation algorithms, which tries to achieve a balance between P-fairness and recommendation accuracy, while also considering user preferences for list diversity (i.e., diversity tolerance). This post-processing step does not depend on any specific recommendation algorithm, and therefore can be widely applied. 

\subsection{Provider-side Fairness}
Fairness for providers (P-fairness) lies in balancing across different providers rather than concentrating on certain dominant ones. Compared to the consumers, the providers are more passive. Providers are not able to seek out recommendation opportunities but have to wait for exposure. Therefore, P-fairness is one of the key issues in MRSs. Take Kiva.org \cite{kiva}, a non-profit site for crowd-sourced microlending, for example. Kiva allows people to make loans via the Internet to low-income entrepreneurs in over 80 countries divided into 8 regions. Kiva has a stated mission of equalizing access to capital across different regions so that loans from each region have the fair probability to be funded. Considering P-fairness gives more chances to the regions that typically receive less attention from lenders. In addition, it also diversifies the recommendation results and allows consumers to explore more new loans that they may be interested in supporting.

\subsection{Diversity Tolerance}
When promoting P-fairness, a slight decline in recommendation accuracy may be inevitable \cite{burke2018balanced}. Instead of being recommended the items that best fit the consumer's need, consumers are also presented with some less optimal items in order to achieve P-fairness. The tolerance towards exploration or diversification varies for different consumers~\cite{eskandanian2017clustering}. Some users may have a strong interest in certain providers while others are more open to seeing items from a variety of providers. Hence, individual users could benefit from personalized ranking strategies, as not all users have equal diversity tolerance. It is preferable that users with high diversity tolerance receive recommended lists that are diversified. 

\subsection{Contributions}
The major contributions of this paper are summarized as follows:
\begin{itemize}
\item We formulate a recommendation scenario in a multi-stakeholder recommender system and define the fairness requirement for providers.
\item We design a re-ranking algorithm to balance between personalization and fairness, and propose the incorporation of diversity tolerance of individuals.
\item We show the results of experiments conducted on synthetic and real-world data to validate the effectiveness of our proposed algorithm. 
\end{itemize}

\section{Fairness-aware Re-ranking}
The fairness-aware problem in MRSs can be naturally formulated as a tradeoff between ranking accuracy and fairness. Given a set of users $U=\{1,\ldots,m\}$, a set of items $V=\{1,\ldots,n\}$ and an initial ranking list $R=[1,\dots,z]$, $z\leq n$. Each item $i\in V$ corresponds to one or more providers $O_i=\{1,\ldots,l\}\subseteq D$, where $D=\{1,\ldots, c\}$ is the set of providers. In other words, each provider $d\in D$ owns a set of items to be recommended. Our task is to generate a re-ranked list $S$ of $K$ ($K\leq z$) distinct items that is both accurate and fair. 

In some contexts, we may identify specific providers as a protected group, whose outcomes are the particular concern. This can be viewed as a special case of our problem where the number of providers is two. In this research, we are looking at more general fairness across all providers and trying to ensure that each re-ranked list $S(u)$ for a specific user $u$ cover as many providers as possible while considering the personalized constraints on diversity tolerance. However, we are not allowed to sacrifice too much accuracy in user preference to achieve an absolute fairness recommendation result. A tradeoff must be made since accuracy and fairness cannot be fully satisfied at the same time \cite{burke_robin_multisided_nodate}.

\subsection{xQuAD}
Result diversification has been studied in the context of information retrieval, especially for web search engines, which have a similar goal to find a ranking of documents that together provide a complete coverage of the aspects underlying a query~\cite{santos2015search}. EXplicit Query Aspect Diversification (xQuAD)~\cite{santos2010exploiting}  explicitly accounts for the various aspects associated with an under-specified query. Items are selected iteratively by estimating how well a given document satisfies an uncovered aspect. However, in the P-fairness problem, simply increasing diversity uniformly in the ranked list for all users may affect user satisfaction. In this work, we proposed a fairness-aware re-ranking on the basis of xQuAD by dynamically adding a personalized bonus to the items of the uncovered providers.

\section{Algorithm}
We build on the xQuAD model to produce a fairness-aware re-ranking algorithm that can be personalized to use tolerance for items from diverse providers. We assume that, for a given user $u$, a ranked recommendation list $R$ has already been generated by a base recommender algorithm. The task of the algorithm is to produce a new re-ranked list $S$ that incorporates provider fairness as well as accuracy. The new list is built iteratively according to the following criterion:


\begin{equation}
P(v|u)+\lambda \tau_u  P(v,\bar S|u),
\label{eq:criterion}
\end{equation}
where $P(v|u)$ is the likelihood of user $u\in U$ being interested in item $v\in V$, predicted by the base recommender. The second term $P(v, \bar S|u)$ denotes the likelihood of user $u$ being interested in an item $v$ not in the currently generated list $S$. Intuitively, the first term incorporates ranking accuracy while the second term promotes provider fairness. The parameter $\lambda$ controls how strongly provider fairness is weighted in general, while $\tau_u$ is a personalized weight learned from each user's historical behavior. When $\tau_u = 1$ for all users, we will refer to this as the Fairness-Aware Re-ranking (FAR) algorithm; when $\tau_u$ is a personalized learned value, we will use the term Personalized Fairness-Aware Re-ranking (PFAR). Both of these variants are compared in the experiments below. The item that scores most highly under the FAR/PFAR metric is added to the output list $S$ and the process is repeated until $S$ has achieved the desired length.


\begin{algorithm}[H]
\caption{(Personalized) Fairness-Aware Re-ranking (FAR/PFAR)}
\begin{algorithmic}[1]
\REQUIRE $u,R,K,\lambda, \tau_u$
\ENSURE $S$
\STATE $S\leftarrow\emptyset$
\WHILE{$|S|< K$} 
\STATE $v^*\leftarrow\arg\max_{v\in R\setminus S}P(v|u)+\lambda\tau_u P(v,\bar S|u)$
\STATE $R\leftarrow R\setminus \{v^*\}$
\STATE $S\leftarrow S\cup\{v^*\}$
\ENDWHILE
\RETURN $S$
\end{algorithmic} 
\label{alg:main}
\end{algorithm}

To achieve P-fairness, the marginal likelihood $P(v, \bar S|u)$ over all providers $d\in D$ is computed by

\begin{equation}
P(v, \bar S|u)  = \sum_{d\in D}P(v, \bar S|d)P(d|u).
\label{eq:marginal}
\end{equation}

Following the approach of~\cite{santos2010exploiting}, we assume that the remaining items are independent of the current contents of $S$ and that the items are independent of each other given the provider $d$. The second assumption may seem rather strong: it amounts to assuming the providers are not ``niche'' providers with specialized inventories. However, our presentation of $P(v|d)$ described below is strictly binary and so having a more nuanced representation of the joint probability of $P(\bar S|d)$ would not change the values computed for the  Eq.\eqref{eq:criterion} criterion. Under these assumptions, we can compute $P(v, \bar S|d)$ in Eq.\eqref{eq:marginal} as
\begin{align}
P(v, \bar S|d)&=P(v|d)P(\bar S|d)\label{subeq: independent}\\
&=P(v|d)\prod_{i\in S}(1-P(i|d))\label{subeq: compensate}.
\end{align}


By substituting Eq.\eqref{subeq: compensate} into Eq.\eqref{eq:marginal}, we can obtain
\begin{equation}
P(v, \bar S|u)=\sum_{d\in D}P(d|u)P(v|d)\prod_{i\in S}(1-P(i|d)).
\label{eq:likelihood}
\end{equation}

We define the likelihood of item $v$ belongs to provider $d$ by
\begin{align*}
    P(v|d)=\begin{cases}
    1 & \text{if }v\in d\\
    0 & \text{if }v\notin d.
    \end{cases}
\end{align*}
This definition is natural since the relationship between items and providers are deterministic and binary. Built on the previous discussion, we can derive the final form of the re-ranking criterion as 

\begin{equation}
P(v|u) + \lambda\tau_u\sum_{d\in D}P(d|u)\mathbbm{1}_{\{v\in d\}}\prod_{i\in S}\mathbbm{1}_{\{i\notin d\}},
\label{eq:indicator}
\end{equation}
where $\mathbbm{1}_{A}$ is the indicator function, having the value 1 if $A$ is true, and 0 otherwise.

The likelihood $P(d|u)$ is the measure of user preference over different providers. This parameter can be tuned according to the importance of a given provider. For example, if the RS would like to promote a specific set of providers, then the weight of these providers could be set higher. By default, we assume a uniform preference over providers. 

The product term $\prod_{i\in S}\mathbbm{1}_{\{i\notin d\}}$ equals to 1 if the current items in $S$ have not covered the provider $d$ yet, and 0 otherwise. In order to select the next item $v^*$ to add to $S$, we compute a re-ranking score for each item in $R\setminus S$ according to Eq.\eqref{eq:indicator}. For an item $v'\in d$, if $S$ does not cover $d$, then an additional positive term will be added to the estimated user preference $P(v'|u)$. Therefore, the chance that $v'$ will be selected is larger, which balances between accuracy and fairness. Note that, as formulated, PFAR favors the items that belong to multiple providers. 

To calculate the user tolerance towards different providers $\tau_u$, we first compute a level of interest $I(d|u)$ of user $u$ for each provider $d\in D$ defined by

\begin{equation}
I(d|u)=\frac{\sum_vr(u,v)\mathbbm{1}_{\{v\in d\}}}{\sum_v\sum_{d'\in D}r(u,v)\mathbbm{1}_{\{v\in d'\}}},
\end{equation}
where $r(u,v)$ is the rating from user $u$ to item $v$.

The preference $I(d|u)\in [0,1]$ indicates the user's taste over providers where $\sum_{d\in D} I(d|u)=1$. Some users may be highly interested in only certain providers, while some users may have equal preferences over all the providers. To capture this characteristic, we use the information entropy \cite{shannon2001mathematical} to identify the user tolerance, namely
\begin{equation}
\tau_u=-\sum_{d\in D}I(d|u)\log I(d|u),
\end{equation}
where a larger $\tau_u$ means the user is more open to a diverse set of providers.

\section{Experiments}
In this section, we test our proposed algorithm on both synthetic data and real-world data. The performance of different re-ranking algorithms are discussed in terms of fairness and accuracy. 

\subsection{Evaluation Metrics}
We aim at providing the ranked list with highest provider coverage rate. To measure this fairness property, we define our first evaluation metric, \textit{average provider coverage rate} (APCR), as follows,

\begin{equation}
    \text{APCR}=\frac{\sum_{u\in U_t}N_{S(u)}}{c|U_t|},
\end{equation}
where $U_t$ is the test user set, and we denote by $|U_t|$ the number of users in the test set. The total number of providers is $c$ and $S(u)$ is the re-ranked list for user $u$. Denote by $N_{S(u)}$  the number of providers covered in the list $S(u)$. A larger APCR indicates a fairer system regarding P-fairness.

\textit{Normalized Discounted Cumulative Gain} (nDCG) is a measure of ranking quality \cite{jarvelin2002cumulated}. We follow the convention in recommender systems that there is a gain in DCG (i.e., the item is \textit{relevant}) if the rating of the item in the list is positive. Since fairness-aware re-ranking aims to achieve a better tradeoff between fairness and accuracy, we expect that there will be a price to pay for obtaining a fairer system. To study the APCR gain under a certain accuracy budget, we calculate $\text{APCR@NDCG}_{5\%}$, the increased APCR value obtained when we allow a $5\%$ decrease in nDCG. We believe that 5\% nDCG loss may be a reasonable accuracy tradeoff if provider fairness can be enhanced. Indeed, it has been shown that user inconsistency in recommender systems generates a lower bound (the ``magic barrier'') within which accuracy measurements are meaningless, so small relaxations of nDCG may not actually represent a loss in performance~\cite{said2012estimating}.

\subsection{Experiments: Synthetic Data}
For our initial experiments, we chose to evaluate the re-ranking method using public data sets that do not have associated provider information, the FilmTrust data set \cite{guo2013novel} and the Movielens 1M data set \cite{movielens}. We created 10 providers and randomly assigned items to each of them, following a geometric distribution, to represent differences among provider inventories.

The FilmTrust data set has 1,508 users and 2,071 items with 35,494 ratings. The Movielens 1M data set contains 6,040 users and 3,706 items with 1,000,209 ratings. For each user, we used a base recommendation algorithm to generate an initial ranked list of 100 items and applied our proposed re-ranking algorithms to select 10 items. We selected four representative base recommenders: 1) itemKNN; 2) userKNN; 3) Alternating Least Squares for Personalized Ranking (rankALS) \cite{takacs2012alternating}; 4) Weighted Regularized Matrix Factorization (WRMF) \cite{hu2008collaborative}. ItemKNN and userKNN are memory-based collaborative algorithms that compute item similarity and user similarity respectively. RankALS directly optimizes the ranking objective function, while WRMF creates a reduced-dimensionality factorization of the rating matrix.

We use the LibRec 2.0 package \cite{guo2015librec} to generate the initial ranking list for all the base recommenders, and all the parameters are tuned on the basis of the default value given in LibRec. We did not spend much effort in parameter tuning since we are interested in the tradeoff between accuracy and fairness of the re-ranking algorithm in this paper, rather than obtaining an optimal nDCG value. The base nDCG and APCR of each recommender are presented in Table \ref{tab:original_result}. On both the FilmTrust and the Movielens data sets, WRMF has the highest nDCG. The APCRs of all the recommenders are around 45\%. All results are averages from five-fold cross validation.

\begin{table}
\centering
\caption{NDCG and APCR of the base recommenders.}
\label{tab:original_result}
\begin{threeparttable}
\begin{tabular}{ccccc}
\toprule
        & \multicolumn{2}{c}{FilmTrust}     & \multicolumn{2}{c}{Movielens}     \\
        & nDCG            & APCR (\%)            & nDCG            & APCR (\%)            \\\midrule
itemKNN & 0.3252          & \textbf{48.26} & 0.0382          & 45.51 \\
userKNN & 0.4856          & 43.30          & 0.1967          & 42.98          \\
rankALS & 0.4052          & 43.39          & 0.1842          & \textbf{45.79}                \\
WRMF    & \textbf{0.5645} & 44.64          & \textbf{0.3317} & 44.37     \\\bottomrule     
\end{tabular}
\end{threeparttable}
\end{table}

In Table~\ref{tab:synthetic}, we report the $\text{APCR@NDCG}_{5\%}$ for FAR and PFAR, compared to the base recommender. With a sacrifice of only 5\% in nDCG, we can gain at least 40\% in provider coverage rate, which validates the effectiveness of our re-ranking method. Our proposed FAR/PFAR methods dynamically balance between accuracy and fairness by adding a fairness score for items belong to the uncovered providers. 

Comparing PFAR to FAR, the $\text{APCR@NDCG}_{5\%}$ value of FAR is generally larger, with an average increase of 12.84\%. This is to be expected since PFAR limits the amount of provider diversification that the re-ranking imposes, based on the individual tolerance, whereas the FAR algorithm treats all users the same, increasing the APCR across the board. Among the four base recommenders, itemKNN has the largest APCR@NDCG$_{5\%}$ for both PFAR and FAR. However, the nDCG for itemKNN is also the lowest: within the context of lower accuracy, it is reasonable to expect higher diversity.


\begin{table*}[t!]
\centering
\caption{$\text{APCR@NDCG}_{5\%}$ on the data sets (\%), providers assigned at random.}
\label{tab:synthetic}
\begin{threeparttable}
\begin{tabular}{c|cccc|cccc}\toprule
        & \multicolumn{4}{c|}{FilmTrust}                       & \multicolumn{4}{c}{Movielens}                       \\
        & \multicolumn{2}{c}{PFAR} & \multicolumn{2}{c|}{FAR} & \multicolumn{2}{c}{PFAR} & \multicolumn{2}{c}{FAR} \\
        & $\lambda$ & APCR@NDCG$_{5\%}$       & $\lambda$ & APCR@NDCG$_{5\%}$        & $\lambda$ & APCR@NDCG$_{5\%}$       & $\lambda$ & APCR@NDCG$_{5\%}$       \\\midrule
itemKNN & 8.79   & 74.92 (+55.25) & 6.94    & 78.90 (+63.49)  & 2.45   & 77.47 (+70.24) & 1.62    & 78.33 (+72.14)  \\
userKNN & 2.52   & 62.63 (+44.63) & 1.66    & 67.86 (+56.72)  & 0.15   & 67.89 (+57.94) & 0.10    & 68.98 (+60.48)  \\
rankALS &    0.08    &       60.99 (+40.54)         &    0.06     &       64.01 (+47.52)          &    0.29    &      72.81 (+59.01)          &    0.22     &     74.43 (+62.57)            \\
WRMF    & 0.81   & 64.57 (+44.66) & 0.49    & 69.17 (+54.96)  & 0.24   & 70.45 (+58.79) & 0.16    & 72.10 (+62.50) \\\bottomrule
\end{tabular}
\end{threeparttable}
\end{table*}

\begin{figure}
\includegraphics[width=\columnwidth]{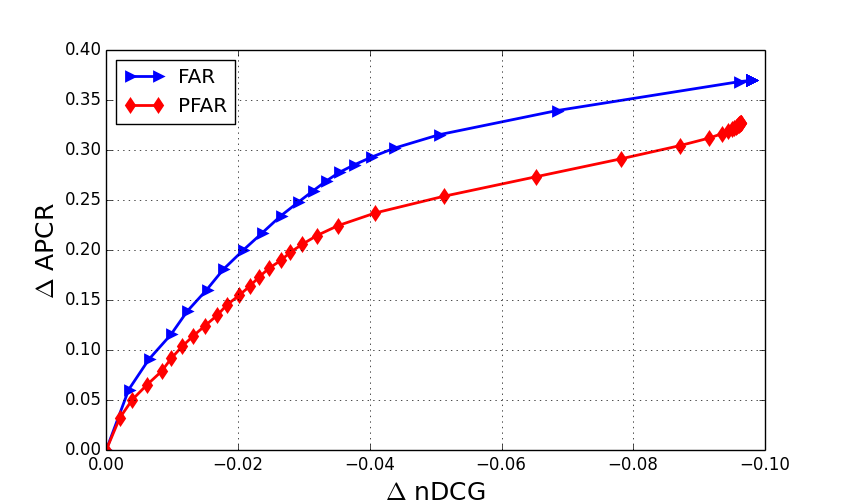}
\caption{Change in nDCG and APCR with increasing $\lambda$ (range 0 to 2.0 in steps of 0.05). FilmTrust data set.}
\label{fig:filmtrust}
\end{figure}

In Figure~\ref{fig:filmtrust}, we focus on the FilmTrust data set and the best-performing WRMF algorithm; the results from the other algorithms and data sets are similar. Change in nDCG and APCR are plotted with increasing values of $\lambda$ from 0 to 2.0 in increments of 0.05. (Base values, as shown in Table~\ref{tab:original_result}, are nDCG=0.5645, APCR=44.64\%.) The increase in APCR and the decline in nDCG show the tradeoff between fairness and accuracy. Throughout the range, FAR dominates PFAR, showing greater APCR improvement at a given degree of nDCG loss.


\begin{figure}
\includegraphics[width=\columnwidth]{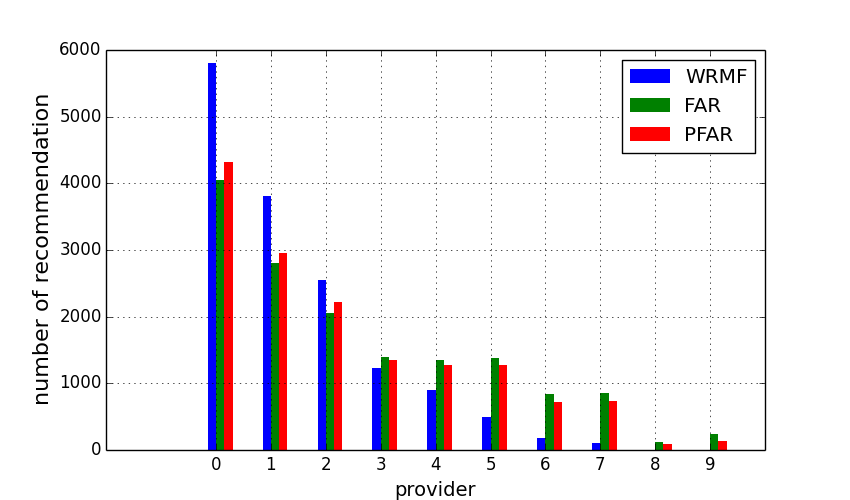}
\caption{Number of recommendations for each provider. Filmtrust data set.}
\label{fig:film_provider}
\end{figure}

To further verify the effectiveness of our proposed algorithm, we compute the number of items being recommended for each provider at a decrease of 5\% in nDCG. These results are shown in Figure~\ref{fig:film_provider} The recommendations generated by the original WRMF focus on a few major providers while paying less attention to the others. FAR/PFAR re-ranking algorithms provide fairer results by promoting the items that belong to less-popular providers. The difference between the algorithms is small, but shows that PFAR is somewhat less balanced than FAR since it considers the different receptivity of the users towards diversified providers. 

\subsection{Experiments: Kiva.org}
We also evaluated the proposed algorithms on a proprietary data set obtained from Kiva.org, including all lending transactions over an 8 month period. As discussed above, Kiva matches lenders with entrepreneurs needing micro-finance. We model the regions that the loans belong to as providers, because part of Kiva.org's mission is to achieve equitable access to capital across regions. As of this writing, Kiva.org does not offer recommendation functionality: in our experiments, we assume a context in which the site provides short lists of recommended loans to users for their support.

The original Kiva.org data set has 853,269 transaction records, involving 113,738 different loans and 178,788 users (lenders). Each loan is specified by features including borrower name, gender, borrower country, loan purpose, funded date, posted date, loan amount, loan sector, and detailed location. One important characteristic of this data set, and the micro-finance domain in general, is there is rapid turn-over in recommendable items. Loans are only available to lenders for a short period until they are fully funded or dropped from the system. Subsequent visitors will not see or be able to support these loans, limiting the maximum item profile size significantly. For example, at a minimum loan amount of \$25, a \$200 loan can have a maximum of only 8 lenders. Contrast this with a consumer taste domain such as MovieLens, where a popular movie might be rated by hundreds or thousands of users. The Kiva.org data set is therefore extremely sparse (with a sparsity of $4.19\times 10^{-5}$) and exhibits a significant item cold-start problem. 

To create a denser data set with greater potential for user profile overlap, we apply a content-based technique, creating \textit{pseudo-items} that represent large categories of items. In particular, we combine into a single pseudo-item all loans that share the same borrower gender, borrower country, loan purpose, loan amount (binned to 5 equal-sized buckets) and loan sector. Then we apply a 10-core transformation, selecting pseudo-items with at least 10 funders and users who had funded at least 10 pseudo-items. The retained data set has 6,410 pseudo-items, 9,999 users and 270,807 ratings, with the sparsity of 0.42\%.

\begin{figure}
\includegraphics[width=\columnwidth]{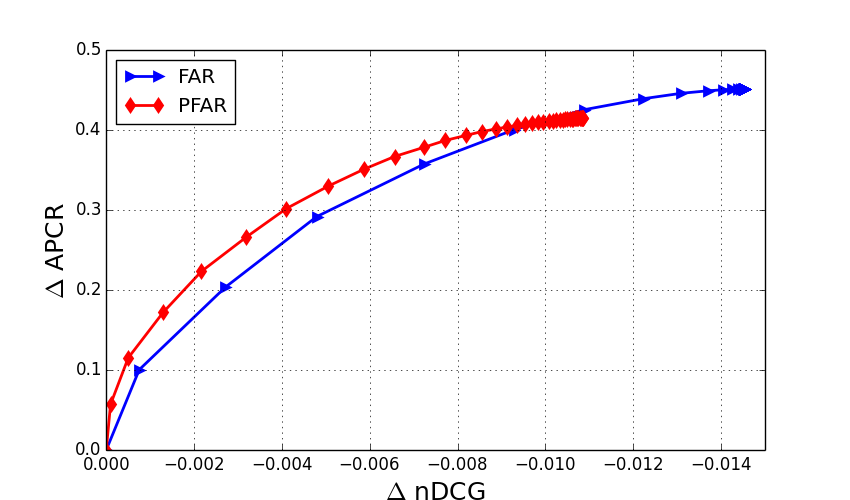}
\caption{Change in nDCG and APCR with increasing $\lambda$ as in Figure~\ref{fig:filmtrust}. Kiva.org data set.}
\label{fig:kiva}
\end{figure}

On this data set, we see a slightly different comparative performance result for PFAR and PFAR. Figure~\ref{fig:kiva} shows the change in nDCG versus the change in APCR for different values of $\lambda$. As in the previous experiment, WRMF was the base recommender and achieved an ACPR of 34.06\% and a nDCG of 0.1026 before re-ranking. With the increase of $\lambda$, APCR increases while nDCG slightly drops. The $\text{APCR@NDCG}_{5\%}$ is 69.16\% (+103.06\%) for PFAR and 69.80\% (+104.97\%) for PFAR. 

Interestingly, in this data set, PFAR does not dominate and actually shows slightly lower APCR for equivalent nDCG loss over most of the $\lambda$ range in our experiment. Here attention to individual differences in diversity tolerance actually enables the system to emphasize a wider range of providers (higher $\lambda$) with a smaller ranking loss.



\begin{figure}
\includegraphics[width=\columnwidth]{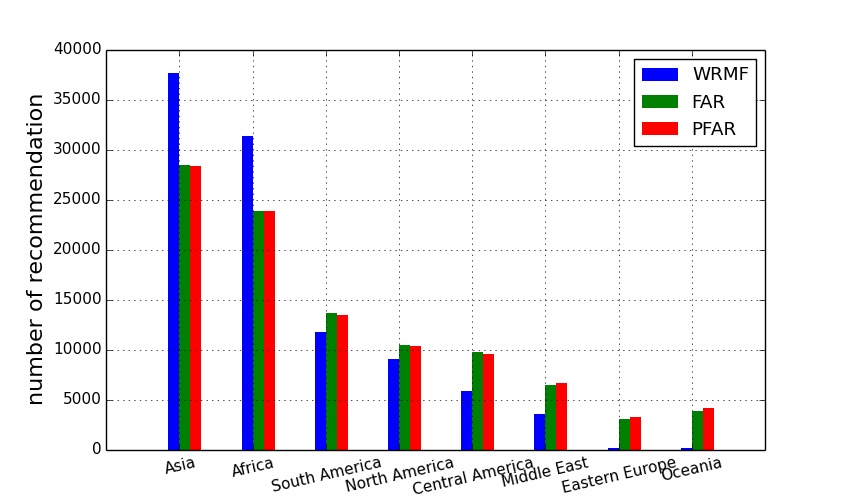}
\caption{Number of recommendations for each region. Kiva.org data set.}
\label{fig:kiva_provider}
\end{figure}

As with FilmTrust, we compute the number of recommendations for each provider when the percentage decrease of nDCG is 5\%, as shown in Figure~\ref{fig:kiva_provider}. For WRMF, most recommendations occurred in Asia and Africa, accounting for 37.66\% and 31.43\% of total recommendations respectively, and Oceania with the smallest percentage at 0.2\%. Table~\ref{tab:inventory} shows the prevalence of each region in the pseudo-item loan inventory, and demonstrates there is certainly over- and under-representation in the WRMF results. Oceania has 10x fewer loans recommended than would be expected if the recommendation lists were proportional to the inventory. For PFAR and PFAR, with a sacrifice of 5\% nDCG, regions like Eastern Europe and Oceania are promoted and the distribution is more balanced.


\begin{table}
\centering
\caption{Inventory of each region in Kiva.org.}
\label{tab:inventory}
\begin{threeparttable}
\begin{tabular}{llr}\toprule
Region          & Inventory  & \% of Total\\\midrule
Asia            & 1619       & 25.3\%  \\
Africa          & 1950       & 30.4\% \\
South America   & 1146       & 17.9\% \\
North America   & 389        & 6.1\% \\
Central America & 629        & 9.8\% \\
Middle East     & 383        & 6.0\% \\
Eastern Europe  & 166        & 2.6\% \\
Oceania         & 128        & 2.0\% \\
\bottomrule
\end{tabular}
\end{threeparttable}
\end{table}

\section{Related Work}

Reciprocal recommendation \cite{reciprocal} is a special case of Multistakeholder Recommender Systems (MRSs), which relies on two-sided interests between two different types of users. For example, RECON \cite{reciprocal} proposes a content-based reciprocal algorithm to learn preferences from both sides of users.

MRSs extend the reciprocal concept to multiple parties instead of two \cite{abdollahpouri_recommender_2017}. Compared to user-centered recommender systems, the fairness problem in MRSs is more challenging since we need to consider the fairness constraints for more than one group of participants \cite{burke_robin_multisided_nodate}. The desired exposure distribution is pre-calculated for each provider for the Behance website in \cite{modani2017fairness}. Fairness is defined by the inverse of the JS-Divergence between the actual exposure distribution and the desired exposure distribution. The balanced neighborhoods method \cite{burke2018balanced} formulates the fairness problem into balancing between protected and unprotected groups by imposing a regularizer on the Sparse Linear Method (SLIM).

In this work, we consider a global fairness for providers (P-fairness) rather than particular protected and unprotected groups. The objective is to let items from different providers have a fair chance of being recommended. This type of P-fairness can be viewed as diversifying recommendation results over providers. In the context of search result diversification, explicit intent-aware diversity frameworks aim at promoting a hybrid of coverage over explicit aspects of a query and novelty. An early example of this kind of algorithm in information retrieval was the maximum marginal relevance (MMR) re-ranking algorithm~\cite{carbonell1998use}. IA-Select \cite{agrawal2009diversifying} maximizes the probability that the average user finds at least one useful result within the top-$K$ results. PM-2 \cite{dang2012diversity} provides search results that best maintains the overall proportionality of the number of documents belonging to different aspects. Intra-list similarity is calculated in \cite{ziegler2005improving} to diversify the topics in the list. xQuAD \cite{santos2010exploiting}, which we build on here, is a probabilistic framework to balance between relevance and novelty. However, result diversification in these information retrieval applications focuses on diversifying the recommendations over topic aspects and does not consider users' varied degrees of diversity tolerance.

The user tolerance towards diversity has been discussed in \cite{eskandanian2017clustering}, where a pre-processing step is designed to cluster users on the basis of their tolerance. The concept of the representative diversity for each category of items is presented in \cite{modani2017fairness}, which is a weighted average of the consumer’s preference for a given category and global preference of that category. The objective function in \cite{modani2017fairness} to be maximized is defined by the multiplication of relevance, fair exposure, and representative diversity. In this paper, we incorporate the user tolerance to control the weight of fairness.

\section{Conclusion and Future Work}
In this work, we designed a personalized fairness-aware re-ranking algorithm for multistakeholder recommender systems (MSRs) that can balance between accuracy and fairness, based on the xQuAD intent-aware diversity enhancement algorithm. We increase the coverage rate of the providers to achieve the fairness for the provider side, and we show that our algorithm can do so with minimal loss in ranking accuracy. In addition, our algorithm includes user-specific weights that can be used to personalize the incorporation of provider diversity based on users' demonstrated tolerance. 

We plan to study a number of variants of the method presented here. We plan to explore different methods for computing personalized diversity tolerance factors, especially to solve the cold-start problem in the current algorithm. We also plan to examine variants of the re-ranking algorithm to take into account the size of each providers' inventory. The xQuAD framework is extremely strict in its requirement that each possible provider appears at least once at the top of the list. Another variant to consider is one that can adjust the accuracy/coverage tradeoff in a dynamic way as items are ranked, valuing accuracy more at the top of the list and provider coverage more at the bottom of the list. Another possibility is to apply a graduated metric of diversity such as MMR. 

Finally, we note that, in real recommendation applications, managing the tradeoff between accuracy and provider coverage is not a single-shot process. Rather it is an online process, where a lack of coverage at one point in time can be compensated for at a later time, and where results are evaluated temporally. This would require making the algorithm sensitive to historical patterns of coverage, rather than just the results obtained in the current list. We intend to explore this type of algorithm design and evaluation in our future work.

%% file: main.bbl

\begin{thebibliography}{24}


\ifx \showCODEN    \undefined \def \showCODEN     #1{\unskip}     \fi
\ifx \showDOI      \undefined \def \showDOI       #1{#1}\fi
\ifx \showISBNx    \undefined \def \showISBNx     #1{\unskip}     \fi
\ifx \showISBNxiii \undefined \def \showISBNxiii  #1{\unskip}     \fi
\ifx \showISSN     \undefined \def \showISSN      #1{\unskip}     \fi
\ifx \showLCCN     \undefined \def \showLCCN      #1{\unskip}     \fi
\ifx \shownote     \undefined \def \shownote      #1{#1}          \fi
\ifx \showarticletitle \undefined \def \showarticletitle #1{#1}   \fi
\ifx \showURL      \undefined \def \showURL       {\relax}        \fi
\providecommand\bibfield[2]{#2}
\providecommand\bibinfo[2]{#2}
\providecommand\natexlab[1]{#1}
\providecommand\showeprint[2][]{arXiv:#2}

\bibitem[\protect\citeauthoryear{Abdollahpouri, Burke, and
  Mobasher}{Abdollahpouri et~al\mbox{.}}{2017a}]%
        {abdollahpouri_recommender_2017}
\bibfield{author}{\bibinfo{person}{Himan Abdollahpouri}, \bibinfo{person}{Robin
  Burke}, {and} \bibinfo{person}{Bamshad Mobasher}.}
  \bibinfo{year}{2017}\natexlab{a}.
\newblock \showarticletitle{Recommender {Systems} as {Multistakeholder}
  {Environments}}. In \bibinfo{booktitle}{\emph{Extended Abstracts of the 25th
  {Conference} on {User} {Modeling}, {Adaptation} and {Personalization}
  ({UMAP}'17)}}. \bibinfo{publisher}{CEUR}, \bibinfo{address}{Bratislava,
  Slovakia}, Article \bibinfo{articleno}{1}, \bibinfo{numpages}{2}~pages.
\newblock



\bibitem[\protect\citeauthoryear{Agrawal, Gollapudi, Halverson, and
  Ieong}{Agrawal et~al\mbox{.}}{2009}]%
        {agrawal2009diversifying}
\bibfield{author}{\bibinfo{person}{Rakesh Agrawal}, \bibinfo{person}{Sreenivas
  Gollapudi}, \bibinfo{person}{Alan Halverson}, {and} \bibinfo{person}{Samuel
  Ieong}.} \bibinfo{year}{2009}\natexlab{}.
\newblock \showarticletitle{Diversifying search results}. In
  \bibinfo{booktitle}{\emph{Proceedings of the second ACM international
  conference on web search and data mining}}. ACM, \bibinfo{pages}{5--14}.
\newblock


\bibitem[\protect\citeauthoryear{Burke, Abdollahpouri, Mobasher, and
  Gupta}{Burke et~al\mbox{.}}{2016}]%
        {soappaper}
\bibfield{author}{\bibinfo{person}{Robin Burke}, \bibinfo{person}{Himan
  Abdollahpouri}, \bibinfo{person}{Bamshad Mobasher}, {and}
  \bibinfo{person}{Trinadh Gupta}.} \bibinfo{year}{2016}\natexlab{}.
\newblock \showarticletitle{Towards Multi-Stakeholder Utility Evaluation of
  Recommender Systems}. In \bibinfo{booktitle}{\emph{UMAP Extended Proceedings:
  Workshop on Surprise, Opposition, and Obstruction in Adaptive and
  Personalized Systems (SOAP 2016)}}. \bibinfo{publisher}{CEUR},
  \bibinfo{address}{Halifax, Canada}.
\newblock


\bibitem[\protect\citeauthoryear{Burke, Sonboli, and Ordonez-Gauger}{Burke
  et~al\mbox{.}}{2018}]%
        {burke2018balanced}
\bibfield{author}{\bibinfo{person}{Robin Burke}, \bibinfo{person}{Nasim
  Sonboli}, {and} \bibinfo{person}{Aldo Ordonez-Gauger}.}
  \bibinfo{year}{2018}\natexlab{}.
\newblock \showarticletitle{Balanced Neighborhoods for Multi-sided Fairness in
  Recommendation}. In \bibinfo{booktitle}{\emph{Conference on Fairness,
  Accountability and Transparency}}. \bibinfo{pages}{202--214}.
\newblock


\bibitem[\protect\citeauthoryear{{Burke, Robin}}{{Burke, Robin}}{2017}]%
        {burke_robin_multisided_nodate}
\bibfield{author}{\bibinfo{person}{{Burke, Robin}}.}
  \bibinfo{year}{2017}\natexlab{}.
\newblock \showarticletitle{Multisided {Fairness} for {Recommendation}}. In
  \bibinfo{booktitle}{\emph{Workshop on {Fairness}, {Accountability} and
  {Transparency} in {Machine} {Learning} ({FATML})}}.
  \bibinfo{publisher}{arXiv}, \bibinfo{address}{Halifax, Nova Scotia}, Article
  \bibinfo{articleno}{arXiv:1707.00093 [cs.CY]}, \bibinfo{numpages}{5}~pages.
\newblock


\bibitem[\protect\citeauthoryear{Carbonell and Goldstein}{Carbonell and
  Goldstein}{1998}]%
        {carbonell1998use}
\bibfield{author}{\bibinfo{person}{Jaime Carbonell} {and} \bibinfo{person}{Jade
  Goldstein}.} \bibinfo{year}{1998}\natexlab{}.
\newblock \showarticletitle{The use of MMR, diversity-based reranking for
  reordering documents and producing summaries}. In
  \bibinfo{booktitle}{\emph{Proceedings of the 21st annual international ACM
  SIGIR conference on Research and development in information retrieval}}. ACM,
  \bibinfo{pages}{335--336}.
\newblock


\bibitem[\protect\citeauthoryear{Dang and Croft}{Dang and Croft}{2012}]%
        {dang2012diversity}
\bibfield{author}{\bibinfo{person}{Van Dang} {and} \bibinfo{person}{W~Bruce
  Croft}.} \bibinfo{year}{2012}\natexlab{}.
\newblock \showarticletitle{Diversity by proportionality: an election-based
  approach to search result diversification}. In
  \bibinfo{booktitle}{\emph{Proceedings of the 35th international ACM SIGIR
  conference on Research and development in information retrieval}}. ACM,
  \bibinfo{pages}{65--74}.
\newblock


\bibitem[\protect\citeauthoryear{Eskandanian, Mobasher, and Burke}{Eskandanian
  et~al\mbox{.}}{2017}]%
        {eskandanian2017clustering}
\bibfield{author}{\bibinfo{person}{Farzad Eskandanian},
  \bibinfo{person}{Bamshad Mobasher}, {and} \bibinfo{person}{Robin Burke}.}
  \bibinfo{year}{2017}\natexlab{}.
\newblock \showarticletitle{A Clustering Approach for Personalizing Diversity
  in Collaborative Recommender Systems}. In
  \bibinfo{booktitle}{\emph{Proceedings of the 25th Conference on User
  Modeling, Adaptation and Personalization}}. ACM, \bibinfo{pages}{280--284}.
\newblock


\bibitem[\protect\citeauthoryear{Guo, Zhang, Sun, and Yorke-Smith}{Guo
  et~al\mbox{.}}{2015}]%
        {guo2015librec}
\bibfield{author}{\bibinfo{person}{Guibing Guo}, \bibinfo{person}{Jie Zhang},
  \bibinfo{person}{Zhu Sun}, {and} \bibinfo{person}{Neil Yorke-Smith}.}
  \bibinfo{year}{2015}\natexlab{}.
\newblock \showarticletitle{LibRec: A Java Library for Recommender Systems.}.
  In \bibinfo{booktitle}{\emph{UMAP Extended Proceedings}}.
  \bibinfo{publisher}{CEUR}, \bibinfo{address}{Dublin, Ireland}, Article
  \bibinfo{articleno}{9}, \bibinfo{numpages}{4}~pages.
\newblock


\bibitem[\protect\citeauthoryear{Guo, Zhang, and Yorke-Smith}{Guo
  et~al\mbox{.}}{2013}]%
        {guo2013novel}
\bibfield{author}{\bibinfo{person}{G. Guo}, \bibinfo{person}{J. Zhang}, {and}
  \bibinfo{person}{N. Yorke-Smith}.} \bibinfo{year}{2013}\natexlab{}.
\newblock \showarticletitle{A Novel Bayesian Similarity Measure for Recommender
  Systems}. In \bibinfo{booktitle}{\emph{Proceedings of the 23rd International
  Joint Conference on Artificial Intelligence (IJCAI)}}.
  \bibinfo{pages}{2619--2625}.
\newblock


\bibitem[\protect\citeauthoryear{Harper and Konstan}{Harper and
  Konstan}{2015}]%
        {movielens}
\bibfield{author}{\bibinfo{person}{F~Maxwell Harper} {and}
  \bibinfo{person}{Joseph~A Konstan}.} \bibinfo{year}{2015}\natexlab{}.
\newblock \showarticletitle{The MovieLens Datasets: History and Context}.
\newblock \bibinfo{journal}{\emph{ACM Transactions on Interactive Intelligent
  Systems (TiiS)}} \bibinfo{volume}{5}, \bibinfo{number}{4}, Article
  \bibinfo{articleno}{19} (\bibinfo{year}{2015}), \bibinfo{numpages}{19}~pages.
\newblock


\bibitem[\protect\citeauthoryear{Hu, Koren, and Volinsky}{Hu
  et~al\mbox{.}}{2008}]%
        {hu2008collaborative}
\bibfield{author}{\bibinfo{person}{Yifan Hu}, \bibinfo{person}{Yehuda Koren},
  {and} \bibinfo{person}{Chris Volinsky}.} \bibinfo{year}{2008}\natexlab{}.
\newblock \showarticletitle{Collaborative filtering for implicit feedback
  datasets}. In \bibinfo{booktitle}{\emph{Data Mining, 2008. ICDM'08. Eighth
  IEEE International Conference on}}. Ieee, \bibinfo{pages}{263--272}.
\newblock


\bibitem[\protect\citeauthoryear{Inc}{Inc}{2018}]%
        {etsy}
\bibfield{author}{\bibinfo{person}{Etsy. Inc}.}
  \bibinfo{year}{2018}\natexlab{}.
\newblock \bibinfo{title}{Etsy.com-Shop for anything from creative people
  everywhere}.
\newblock
\newblock
\urldef\tempurl%
\url{https://www.etsy.com/}
\showURL{%
\tempurl}


\bibitem[\protect\citeauthoryear{Jannach, Zanker, Felfernig, and
  Friedrich}{Jannach et~al\mbox{.}}{2010}]%
        {jannach2010recommender}
\bibfield{author}{\bibinfo{person}{D. Jannach}, \bibinfo{person}{M. Zanker},
  \bibinfo{person}{A. Felfernig}, {and} \bibinfo{person}{G. Friedrich}.}
  \bibinfo{year}{2010}\natexlab{}.
\newblock \bibinfo{booktitle}{\emph{Recommender systems: an introduction}}.
\newblock \bibinfo{publisher}{Cambridge University Press}.
\newblock


\bibitem[\protect\citeauthoryear{J{\"a}rvelin and
  Kek{\"a}l{\"a}inen}{J{\"a}rvelin and Kek{\"a}l{\"a}inen}{2002}]%
        {jarvelin2002cumulated}
\bibfield{author}{\bibinfo{person}{Kalervo J{\"a}rvelin} {and}
  \bibinfo{person}{Jaana Kek{\"a}l{\"a}inen}.} \bibinfo{year}{2002}\natexlab{}.
\newblock \showarticletitle{Cumulated gain-based evaluation of IR techniques}.
\newblock \bibinfo{journal}{\emph{ACM Transactions on Information Systems
  (TOIS)}} \bibinfo{volume}{20}, \bibinfo{number}{4} (\bibinfo{year}{2002}),
  \bibinfo{pages}{422--446}.
\newblock


\bibitem[\protect\citeauthoryear{kiva}{kiva}{2018}]%
        {kiva}
\bibfield{author}{\bibinfo{person}{kiva}.} \bibinfo{year}{2018}\natexlab{}.
\newblock \bibinfo{title}{Kiva - Loans that change lives}.
\newblock
\newblock
\urldef\tempurl%
\url{https://www.kiva.org/}
\showURL{%
\tempurl}


\bibitem[\protect\citeauthoryear{Modani, Jain, Soni, Gupta, and Agarwal}{Modani
  et~al\mbox{.}}{2017}]%
        {modani2017fairness}
\bibfield{author}{\bibinfo{person}{Natwar Modani}, \bibinfo{person}{Deepali
  Jain}, \bibinfo{person}{Ujjawal Soni}, \bibinfo{person}{Gaurav~Kumar Gupta},
  {and} \bibinfo{person}{Palak Agarwal}.} \bibinfo{year}{2017}\natexlab{}.
\newblock \showarticletitle{Fairness Aware Recommendations on Behance}. In
  \bibinfo{booktitle}{\emph{Pacific-Asia Conference on Knowledge Discovery and
  Data Mining}}. Springer, \bibinfo{pages}{144--155}.
\newblock


\bibitem[\protect\citeauthoryear{Pizzato, Rej, Chung, Koprinska, and
  Kay}{Pizzato et~al\mbox{.}}{2010}]%
        {reciprocal}
\bibfield{author}{\bibinfo{person}{Luiz Pizzato}, \bibinfo{person}{Tomek Rej},
  \bibinfo{person}{Thomas Chung}, \bibinfo{person}{Irena Koprinska}, {and}
  \bibinfo{person}{Judy Kay}.} \bibinfo{year}{2010}\natexlab{}.
\newblock \showarticletitle{RECON: a reciprocal recommender for online dating}.
  In \bibinfo{booktitle}{\emph{Proceedings of the fourth ACM conference on
  Recommender systems}}. \bibinfo{publisher}{ACM}, \bibinfo{address}{New York,
  NY, USA}, \bibinfo{pages}{207--214}.
\newblock


\bibitem[\protect\citeauthoryear{Said, Jain, Narr, Plumbaum, Albayrak, and
  Scheel}{Said et~al\mbox{.}}{2012}]%
        {said2012estimating}
\bibfield{author}{\bibinfo{person}{Alan Said}, \bibinfo{person}{Brijnesh~J
  Jain}, \bibinfo{person}{Sascha Narr}, \bibinfo{person}{Till Plumbaum},
  \bibinfo{person}{Sahin Albayrak}, {and} \bibinfo{person}{Christian Scheel}.}
  \bibinfo{year}{2012}\natexlab{}.
\newblock \showarticletitle{Estimating the magic barrier of recommender
  systems: a user study}. In \bibinfo{booktitle}{\emph{Proceedings of the 35th
  international ACM SIGIR conference on Research and development in information
  retrieval}}. ACM, \bibinfo{pages}{1061--1062}.
\newblock


\bibitem[\protect\citeauthoryear{Santos, Macdonald, and Ounis}{Santos
  et~al\mbox{.}}{2010}]%
        {santos2010exploiting}
\bibfield{author}{\bibinfo{person}{Rodrygo~LT Santos}, \bibinfo{person}{Craig
  Macdonald}, {and} \bibinfo{person}{Iadh Ounis}.}
  \bibinfo{year}{2010}\natexlab{}.
\newblock \showarticletitle{Exploiting query reformulations for web search
  result diversification}. In \bibinfo{booktitle}{\emph{Proceedings of the 19th
  international conference on World wide web}}. ACM, \bibinfo{pages}{881--890}.
\newblock


\bibitem[\protect\citeauthoryear{Santos, Macdonald, Ounis,
  et~al\mbox{.}}{Santos et~al\mbox{.}}{2015}]%
        {santos2015search}
\bibfield{author}{\bibinfo{person}{Rodrygo~LT Santos}, \bibinfo{person}{Craig
  Macdonald}, \bibinfo{person}{Iadh Ounis}, {et~al\mbox{.}}}
  \bibinfo{year}{2015}\natexlab{}.
\newblock \showarticletitle{Search result diversification}.
\newblock \bibinfo{journal}{\emph{Foundations and Trends{\textregistered} in
  Information Retrieval}} \bibinfo{volume}{9}, \bibinfo{number}{1}
  (\bibinfo{year}{2015}), \bibinfo{pages}{1--90}.
\newblock


\bibitem[\protect\citeauthoryear{Shannon}{Shannon}{2001}]%
        {shannon2001mathematical}
\bibfield{author}{\bibinfo{person}{Claude~Elwood Shannon}.}
  \bibinfo{year}{2001}\natexlab{}.
\newblock \showarticletitle{A mathematical theory of communication}.
\newblock \bibinfo{journal}{\emph{ACM SIGMOBILE mobile computing and
  communications review}} \bibinfo{volume}{5}, \bibinfo{number}{1}
  (\bibinfo{year}{2001}), \bibinfo{pages}{3--55}.
\newblock


\bibitem[\protect\citeauthoryear{Tak{\'a}cs and Tikk}{Tak{\'a}cs and
  Tikk}{2012}]%
        {takacs2012alternating}
\bibfield{author}{\bibinfo{person}{G{\'a}bor Tak{\'a}cs} {and}
  \bibinfo{person}{Domonkos Tikk}.} \bibinfo{year}{2012}\natexlab{}.
\newblock \showarticletitle{Alternating least squares for personalized
  ranking}. In \bibinfo{booktitle}{\emph{Proceedings of the sixth ACM
  conference on Recommender systems}}. ACM, \bibinfo{pages}{83--90}.
\newblock


\bibitem[\protect\citeauthoryear{Ziegler, McNee, Konstan, and Lausen}{Ziegler
  et~al\mbox{.}}{2005}]%
        {ziegler2005improving}
\bibfield{author}{\bibinfo{person}{Cai-Nicolas Ziegler},
  \bibinfo{person}{Sean~M McNee}, \bibinfo{person}{Joseph~A Konstan}, {and}
  \bibinfo{person}{Georg Lausen}.} \bibinfo{year}{2005}\natexlab{}.
\newblock \showarticletitle{Improving recommendation lists through topic
  diversification}. In \bibinfo{booktitle}{\emph{Proceedings of the 14th
  international conference on World Wide Web}}. ACM, \bibinfo{pages}{22--32}.
\newblock


\end{thebibliography}
